Version Dec. 14, 2008

# Vortex dynamics in confined stratified conditions

Farkhad G. Aliev[1], Juan F. Sierra[1], Ahmad A. Awad[1], Gleb N. Kakazei[2], Dong-Soo Han[3], Sang-Koog Kim[3], Vitali Metlushko[4], Bojan Ilic[5] and Konstantin Y. Guslienko[3]

[1] *Dpto. Fisica de la Materia Condensada, CIII, Universidad Autonoma de Madrid, Spain*

[2] *IFIMUP-IN, Departamento de Fisica, Universidade do Porto, Porto, Portugal*

[3] *Research Center for Spin Dynamics & Spin-Wave Devices, Seoul National University, Seoul, South Korea*

[4] *Dept. Electrical and Computer Engineering, University of Illinois at Chicago, Chicago, Illinois, USA*

[5] *Cornell Nanofabrication Facility, Cornell University, Ithaca, USA*

We report on linear spin dynamics in the vortex state of the Permalloy dots subjected to stratified (magnetic) field. We demonstrate experimentally and by simulations the existence of *two distinct dynamic regimes corresponding to the vortex stable and metastable states*. Breaking cylindrical symmetry leads to unexpected eigenmodes frequency splitting in the *stable* state and appearance of new eigenmodes in the *metastable* state above the vortex nucleation field. Dynamic response in the *metastable* state strongly depends on relative orientation of the external rf pumping and the bias magnetic fields. These findings may be relevant for different vortex states in confined and stratified conditions.

In many physical systems such as liquids, plasma, superconductors, etc. the topological excitations (vortices) are frequently subjected to stratified conditions [1-4]. In practice such vortices also often satisfy some confinement conditions due to finite sizes of the systems. The strong pinning boundary conditions for dynamic magnetization in thin magnetic particles (dots) at the lateral system edges [5] are the simplest confinement conditions. Special interest in the magnetic vortices is inspired by the possibility of easy dynamical switching of the vortex core magnetization direction [6] that has been suggested as a potentially new root to create nanoscale memory units for data storage. Fundamental studies have been stimulated by lack of understanding of the spin wave dynamics in the vortex-state dots presenting the simplest possible nanosystem with topological non-uniform magnetization. Precise mapping of the high frequency spin excitation eigenmodes, *especially the eigenmodes breaking axial symmetry*, is of great importance because they are modes expected to define the vortex switching characteristic times. Here we consider the dynamic eigenmodes excited by parallel or perpendicular microwave field in a circular magnetic dot in the vortex state displaced by the external in-plane magnetic field.

It is known that without the bias field multiple spin eigenmodes in the magnetic vortex state can be excited by an external perturbation and the dynamical response in the linear regime is described by their superposition. The lowest frequency excitation corresponds to the gyrotropic mode when the vortex moves as a whole around an equilibrium position [7-9], meanwhile the higher frequency modes correspond to the spin waves excited mainly outside of the vortex core [10-19]. The spin wave having radial or azimuthal symmetry with respect to the dot center are described by integers ($n, m$), which indicate number of nodes in the dynamic magnetization along radial ($n$) and azimuthal ($m$) directions. For in-plane magnetic excitation field the azimuthal spin waves $m = \pm 1$ and the gyrotropic mode are expected to be excited [13,16,17] because only these spin modes have non zero average in-plane magnetization and can interact with the uniform rf driving field. Previous studies were mainly focused on resonant response in the absence of bias magnetic field [6-9,11,12,14,15] when the static vortex was localized in the dot center. Our letter reports on broadband measurements of the vortex dynamics in circular Permalloy (Py) dots excited by applying in-plane stratifying field with the variable angle between the bias and excitation fields. We identify new unexpected eigenmodes both in the stable and metastable vortex states and show that the eigenmodes excited in the metastable state above the nucleation field depend strongly on the relative orientation of the driving and stratifying fields. Our experimental results and simulations could be relevant for different vortex systems in the confined stratified conditions.

Two sets of square arrays of Py circular dots were fabricated by combination of lithography and lift-off techniques on a standard Si(100) substrate as explained elsewhere [15,20,21]. The first set includes three samples with thickness $L$=50 nm, diameter $D$=1000 nm and the dot´s lattice parameters (center-to-center distance, $d$) of 1200, 1500 and 2500 nm. The second set included two arrays of Py dots with thickness $L$=25 nm and diameters of $D$=1035 nm ($d$=2000 nm) and of 572 nm ($d$=1000 nm). The excited spin waves have been studied at room temperature by broadband spectrometer based on vector network analyzer [22]. The set-up allows to apply rf field either in perpendicular configuration ($\mathbf{h}_{rf\perp}$) where $\mathbf{H}_{bias} || \mathbf{x}$ and $\mathbf{h}_{rf} \perp \mathbf{x}$ as shown in Fig.1, or with parallel pumping scheme ($\mathbf{h}_{rf||}$) when $\mathbf{H}_{bias} || \mathbf{x}$ and coplanar wave guide (CPW) rotated perpendicularly with respect to the in-plane static bias field, providing $\mathbf{h}_{rf} || \mathbf{x}$. The data were analyzed on the basis of transmission model developed under the assumption that the dominant CPW mode is the TEM mode and also neglecting the effect of reflection [16]. The estimated magnitude of in-plane rf field is below 0.2 Oe.

Besides the experimental studies, the micromagnetic simulations [23] were carried out for a circular shaped Py dot having the thickness of 25 nm and the diameter of 1035 nm. The physical parameters of the individual cells of $5\times5\times25$ nm$^3$ used were: the exchange stiffness constant $A = 1.4 \times 10^{-11}$ (J/m), the saturation magnetization $M_s = 830\times10^3$ A/m, $\gamma/2\pi = 2.96$ MHz/Oe taken from the measurements [20], and the Gilbert damping constant of $\alpha = 0.01$.

We first identify the vortex nucleation ($H_n$) and annihilation ($H_a$) fields for the dots having the vortex remanent state. These fields describe a range of field stability of the vortex ground state. Figure 2a shows the typical dependence of the dot array magnetization (*M*) on the in-plane magnetic field during the hysteresis cycle for the positive field branch with $H_n$ and $H_a$ fields marked by vertical arrows. In order to compare the static and dynamic measurements we normalized the external field by $H_a$.

To study the vortex ground state excitations we first saturated *M* by the in-plane magnetic field (*H*) above the annihilation field ($H>3H_a$). Then, the magnetic field was swept within the interval $-3H_a<H<3H_a$ creating and annihilating the vortex state. We present here the experiments and simulations of spin dynamics in Py dots with thickness of 25 nm and diameter of 1035 nm by using either parallel (**h**$_{rf||}$, Fig.2b,d) or perpendicular (**h**$_{rf\perp}$, Fig. 2c,e) rf drive. The similar results have been also obtained for other dot arrays with the aspect ratio thickness/radius (*L/R*) varied between 0.05 and 0.1 revealing extra higher frequency modes appearing in the measured spectral window for smaller *L/R*. The dynamic response remains qualitatively unaffected by the interdot dipole-dipole interaction with more than 2-fold change in the interdot distance ensuring that we are observing a single dot eigenmodes.

Let us first discuss the experimental results shown in Fig. 2. Applying the uniform rf field **h$_{rf}$** we excite only the spin eigenmodes localized in the areas where the torque **h** x **M**$_0 \neq$ 0, (**M**$_0$ is the static magnetization). Increasing *H* reveals three main field regions in the excitation spectra: (i) only single vortex is stable (SV); the quasi-uniform and vortex states are stable simultaneously (metastable vortex, MV); and the quasi-uniform or saturated state (US). In the SV regime ($H<H_n$) two doublets of the spin eigenmodes are observed with the eigenfrequencies being independent of the orientation of the rf field (see Fig. 2b,c). These low-field doublets [24] can be described as azimuthal spin waves with the indices $m=\pm 1$, $n=0$, indicated in Fig.2 as (1), and with the indices $m=\pm 1$, $n=1$. The later modes have much lower intensity and will not be discussed further. The zero-field splitting of the spin eigenfrequencies of the doublets is about ~1GHz in accordance with recently developed model [24] of the dynamical vortex core – spin wave interaction.

The eigenmodes classification is strictly applicable to the zero-field case, but it can be approximately used up to disappearing of the azimuthal modes at $H \approx H_n$. At small $H$ one can excite the eigenmodes with indices $m=0, \pm 2$, intensities of which are proportional to $H^2$. The first mode is not observed in our simulations (see below), whereas the $m=\pm 2$ modes are responsible for formation of the spin wave branch (2) observed with parallel pumping. The "soft" mode (3) exists also only with parallel pumping when vortex core is close to the dot edge at $H_n < H < H_a$ and disappears at $H \geq H_a$. Increasing the bias field with the perpendicular pumping also suppresses the azimuthal modes $|m|=1$ at $H>H_n$. The lowest zero-field mode ($m=+1$, $n=0$), however, reveals an additional splitting at low fields for both the pumping schemes. This frequency splitting is unexpected because this mode is not degenerate.

It is important to mention that *the modes observed in the MV state ($H_n<H<H_a$) are qualitatively different when excited with perpendicular or parallel pumping*. Indeed, a strongly field dependent mode (3) is observed for the MV with the $\mathbf{h}_{rf\parallel}$ configuration (see Fig.2b,d). The frequency of this eigenmode decreases with increasing magnetic field. In contrast, in the $\mathbf{h}_{rf\perp}$ scheme a strong parabolic-like mode marked as (3´) is excited in the MV state accompanied by multiple satellites (Fig.2c,e). Moreover, near the annihilation field $H_a$ the most contrast eigenmode (3´) transforms abruptly into the almost uniform precession mode (4) existing at $H>H_a$ in the US. This quasi-uniform mode can be excited also by the parallel pumping (Fig.2b) due to non-perfect field alignment.

To understand the excited spin eigenmodes in the frequency domain at given external static field we conducted the micromagnetic simulations by applying the bias field to the vortex-state dot along the *x*-direction in the range from 0 to 1000 Oe with the steps of 20 Oe. To excite the vortex ground state, similarly to the experiment, we applied the variable magnetic field along two different in-plane directions, parallel and perpendicular to the bias field. Both the $h_{rf||}$ and $h_{rf\perp}$ schemes were explored by applying sine-field of variable frequency and with the amplitude of 1 Oe. Figures 2 d,e show the simulated Fourier power spectra at given range of frequencies as a function of external field.

Figure 3 shows simulated spatial distributions of the dynamic magnetization for the parallel ($\Delta M_x/M_s$) and the perpendicular ($\Delta M_y/M_s$) pumping schemes for the most intensive excited modes (marked as 1-4 in Fig.2) at some specific frequencies and normalised fields. The lowest ($n=0$, $m=+1$) azimuthal mode (1) is observed in zero bias field. The application of a weak field (e.g., $H/H_a=0.077$) smoothly breaks the vortex state cylindrical symmetry but maintains the character of the mode (1). As pointed above, the mode (2) splitting out from the lowest azimuthal mode *has qualitatively new character* describing approximately as a superposition of the azimuthal spin wave modes $m=\pm 2$ rotating around the vortex-core in the opposite directions (clockwise, CW and counter-clockwise, CCW). Given the vortex core shift in the positive *y*-direction, the modes start from the positive *y* semi-axis, propagate, meet each other and disappear at negative *y* semi-axis. When the stratifying field is further increased (4.86GHz, $H/H_a=0.423$), the mode (2) also shows similar qualitatively new motion but with more nodes in disturbed azimuthal direction (Fig.3). We can distinguish two different regions in the dynamical *M*-images. For the first one, close to vortex core, there is a superposition of the CCW and CW motions. In the second region (far from the vortex core, the lower part of the dot) *standing-like spin wave modes dominate* (Fig.3, mode (2), $f$ =4.86 GHz). The increase of the bias field expands the second region. On the other hand, there is a clear standing modes dominating pattern for the lower frequency mode (mode (3), 4.18GHz, MV) with reduced area occupied by the azimuthal-like modes. This mode (3) can be approximately described as having a wave vector along the bias field (analogy to the backward volume magnetostatic spin waves in continuous films) because the nodal planes are perpendicular to the *x*-direction.

When we applied the perpendicular drive, the response in the low field region (SV) is the same as in the parallel case. In the MV region the high frequency mode (3′) is more intensively excited compared to the parallel pumping case (the mode (3)). To examine the modes we conducted simulations with the frequency and bias field as follows: (5.84GHz, 0.423), (6.88GHz, 0.923). In general, the dynamic *M*-images are similar to those observed for the mode (3) for weaker fields in the MV state, and become more complicated close to boundary with the US. We note that transition to the US regime suppresses abruptly both the (3′) and (3) modes. For complete sequences of the data shown in Fig.3, see movies in Ref. [25]. We attribute small (<10%) disagreement between the experiment and simulated eigenmodes to (i) influence of the interdot dipolar interaction; (ii) presence of small out-of-plane rf component (Fig. 1) and (iii) thin (about 2 nm) Py oxide layer [26], which influences the dot thickness and therefore the spin eigenfrequencies [10].

*In conclusion,* breaking cylindrical symmetry of the ground state for confined vortex leads to unexpected eigenmodes frequency splitting in the stable vortex state and appearance of new eigenmodes excited in the metastable state. It is shown that the vortex dynamics in confined stratified conditions strongly depends on the relative orientation of the driving and bias fields. The vortex state of the magnetic dots excited in different directions with respect to the bias field can provide an unique information regarding dynamics in confined stratified conditions in other vortex systems, *e.g.*, in the arrays of the vortex nano-oscillators excited by spin-polarized current [27].

# Figure captions

**Figure 1**

Sketch of the experimental setup and the coordinate system used. Insert shows the typical SEM image of an array of Py dots. The bias magnetic field is directed along *x* or *y* axes.

**Figure 2**

(a) Magnetization vs. magnetic field normalized by annihilation field ($H_a$ = 360 Oe). Arrows indicate the vortex nucleation and annihilations fields.

(b-e) Field dependence of the excitation frequencies of Permalloy dots. Intensity plots of the measured and simulated spectra for the Py dot arrays with thickness 25 nm and 1035 nm diameter for rf field parallel (b,d) and perpendicular (c,e) to the bias magnetic field. Stable vortex state (SV), metastable vortex state (MV) and quasi-uniform magnetic state (US) are separated by the vertical lines corresponding to the vortex nucleation and annihilations fields. Numbers 1, 2, 3, 3´ and 4 are modes discussed in the text.

**Figure 3**

Simulated (in the form of time sequences with the step of about ¼ oscillation period) spatial distributions of the vortex dynamic magnetization in the SV and MV field regions. Numbers 1, 2, 3 denote the frequency branches for parallel pumping (the magnetization component $\Delta M_x/M_s$) and 3´ for perpendicular pumping (the component $\Delta M_y/M_s$) in Fig. 2. The first number in the brackets ($f$, $H/H_a$) shows the eigenmode frequency, whereas the second number shows the normalized magnetic bias field. We selected the frequency/field points, where splitting of the eigenfrequencies occurs at each frequency branch.

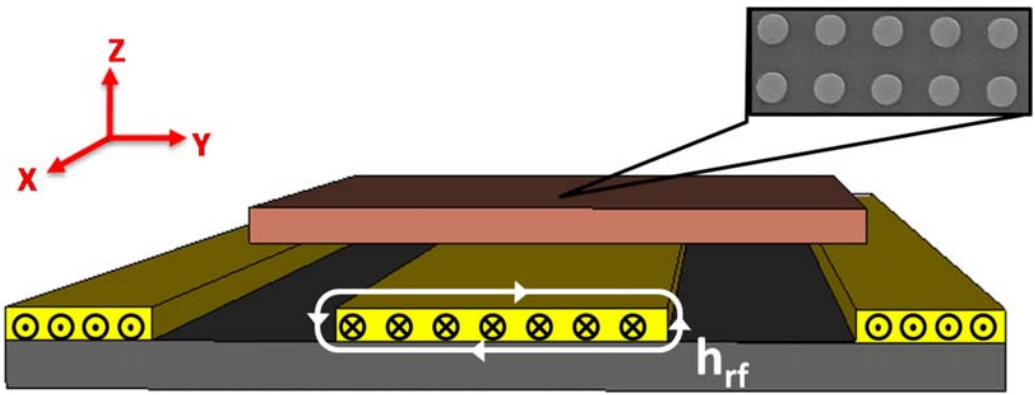

**Figure 1**

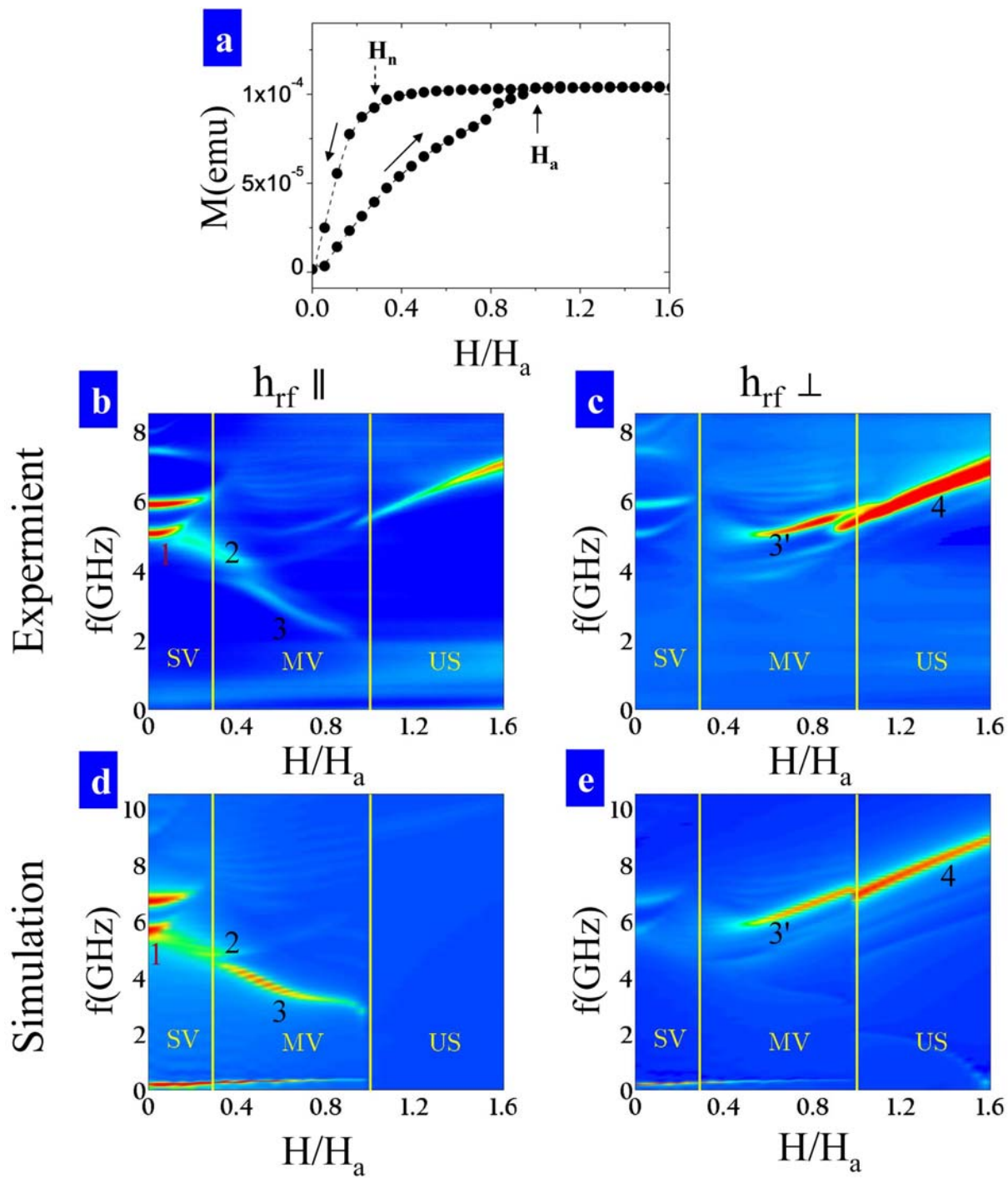

**Figure 2**

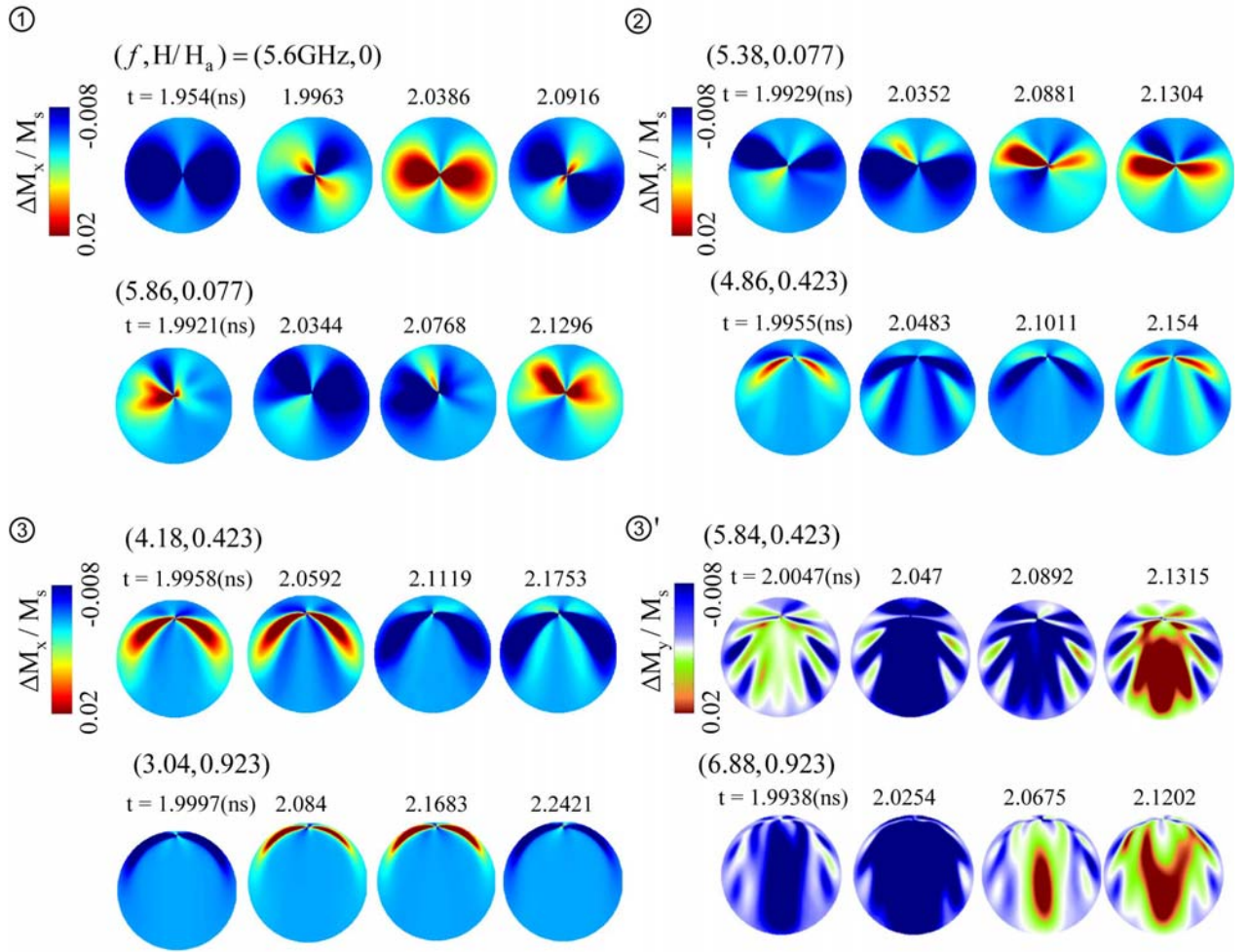

**Figure 3**